\documentclass[fleqn,twoside]{article}
\usepackage{espcrc2}

\usepackage{graphicx}
\usepackage[figuresright]{rotating}

\newcommand{\AmS}{{\protect\the\textfont2
  A\kern-.1667em\lower.5ex\hbox{M}\kern-.125emS}}

\title{Concluding Remarks/Summary}

\author{Sandip Pakvasa\address{Department of Physics and Astronomy \\
            University of Hawaii, Honolulu, HI  96822}} 
\begin{document}

\begin{abstract}
These remarks summarize some of the discussion at the Now 2004; 
in addition some topics not touched on at the meeting are 
reviewed briefly. 
\end{abstract}

\maketitle

\section{Introduction}
Sometimes one hears the lament\cite{lederman} that while theorists are
lionized, rarely do experimenters get the credit that they deserve.  To
redress this balance a little, I would like to begin by recalling the
life and times of Charles Drummond Ellis; whose work made 
the invention of the neutrino inevitable.  His romantic story starts with his
``conversion'' from an army career as an artillery officer to research in
physics while interned in  prison in Berlin during the first world war.  
During the period
1921-27, he collaborated with James Chadwick and William Wooster in a series of
measurements of $\beta$-ray spectra which established beyond doubt that
the spectra were continuous\cite{ellis} and ended the on-going controversy
with Lise Meitner.  In my book Ellis is one of the first heroes in the
neutrino story\cite{full}.

Recalling NOW 2000\cite{proceedings}, the progress in the 
last four years has been
impressive, even spectacular. Then, both SNO and Kamland were still in the future, and now we have
results from them.  Similarly, K2K results have also been
published.  Oscillation dips have been observed in atmospheric neutrinos
by Super-K and in reactor anti-neutrinos by Kamland.  Several new
facilities are close to completion.

\section{Terra Cognita I}
With the results of SNO and Kamland combined with all the previous results,
one can summarize the current knowledge\cite{valle,petcov} 
of the neutrino mass differences
and mixings:
\vspace{-1mm}
\begin{eqnarray}
\left (
\begin{array}{c}
\nu_e \\ 
\nu_\mu \\ 
\nu_\tau
\end{array} \right )
= U_{MNSP} 
\left (
\begin{array}{c}
\nu_1 \\
\nu_2 \\
\nu_3
\end{array} \right )
\end{eqnarray}
where
\begin{equation}
U_{MNSP} = U \ \cong
\left  ( 
\begin{array}{ccc}
c_{12} & -s_{12} & U_{e3} \\
s_{12}/\sqrt{2} & c_{12}/\sqrt{2} & -1/\sqrt{2}\\
s_{12}/\sqrt{2} & c_{12}/\sqrt{2} & 1/\sqrt{2}\\ 
\end{array} \right )
\end{equation}
In the above, the smallness of $\theta_{13}$, and the closeness of 
$\theta_{23}$ to  $45^0$ have been incorporated; the actual fits correspond to
 $\theta_{12} \cong 33 \pm 2^0$, $\theta_{23} 
\cong 45 \pm 11^0,|U_{e3}| < 0.15, 
|\delta m^2_{32}|\simeq2.5.10^{-3} eV^2$ and
$\delta m^2_{21} = m^2_2 - m_1^2 = + 8.10^{-5} eV^2.$

We know $m_2 > m_1$ for the MSW effect to be operative in the sun, but the
location of $m_3$ w.r.t. $m_2$ in unknown.  Likewise we do not know
the offset of the three masses from zero.

The above simple picture is muddied by the LSND ``effect'', where in
$\pi^+$ decay at rest, $\pi^+ \rightarrow \mu^+ \nu_\mu, \ \mu^+
\rightarrow e^+ \nu_e \bar{\nu}_\mu$, the wrong neutrino $\bar {\nu}_e$
was seen at a fractional rate of about (2-3) $10^{-3}$. The simplest
extension of three neutrinos to include one more sterile $\nu_s$ is now
constrained strongly by the totality of neutrino data and disfavored.  
Two sterile (in addition to the three flavor) neutrinos provides a
consistent fit to all data so far\cite{shaeritz}.  
One may ask why two light sterile
neutrinos, but then why not?  Sterile neutrinos were avoided altogether by
allowing CPT violation\cite{murayama}, wherein 
$\delta m^2_{solar}$ and $\delta m^2_{KL}$
would be different.  However, there is no indication in data for such a
difference.  A baroque solution would be to violate CPT {\it and} have one
light sterile neutrino (or have decoherence)\cite{barger}.  
A proposal was made\cite{babu}, based on a rare decay mode of $\mu^+:
\mu^+ \rightarrow e^+ \bar{\nu}_e \bar{\nu}_\alpha$, violating lepton
number by 2, with a fractional rate of $\sim 3.10^{-3}$,
which would not involve neutrino oscillations.  But this too is now ruled
out by TWIST results\cite{gaponenko} on the precision 
measurement of the Michel parameter
in $\mu$-decay, and by the limits from KARMEN\cite{karmen}.  We now must await results from MINI-BOONE on $\nu_\mu
\rightarrow \nu_e$ oscillations to be announced a year from now (summer
2005?). Eventually MINI-BOONE\cite{shaeritz} might also measure
$\bar{\nu}_\mu-\bar{\nu}_e$ oscillations.  If the LSND result is confirmed,
there can be a significant (and positive) impact on the expected size of
any CPV effect in neutrino oscillations\cite{barger1}.

In 2000, non-oscillatory explanations were still possible for both solar
and atmospheric neutrino data.  For example, atmospheric neutrino data
could be explained by neutrino decay, decoherence or by Flavor Changing 
Neutral Currents(FCNC).  
The solar neutrino data could be accounted for by FCNC, Resonant Spin-Flip Precession(RSFP) or Lorentz Invariance Violation(LV).  We
have learned that all of these possibilities are now excluded\cite{alan}, except
as sub-leading effects.  We can now probe non-standard interactions(NSI)
of neutrinos in solar neutrino data as well atmospheric neutrino data.  
When NSI are allowed as sub-leading effects, the oscillation fits get
modified; resulting in significant deviations in the fitted parameters.  
For example, in solar fits, the so-called dark-side solutions $(i.e. \ m_1 >
m_2)$ can emerge\cite{valle}.

\subsection{Solar and Reactor Neutrinos}
  It seems that we have to wait for a few
months for the full detailed paper on SNO results of the salt phase and a
year or more for results of the run with $^3He$ counters\cite{chen}.  
It would be of
great interest to look for (and find) the expected upturn in the energy
spectrum at about 6 MeV and the expected level of the day-night asymmetry
of about 2.5\%.  These would be powerful confirmations of LMA solution and
the MSW effect at work.

As for reactor $\bar{\nu}_e's$ in Kamland, we heard about the improvements
in statistics, energy calibration and in fiducial volume.  We also learned
about the new background from $^{210}Pb$ which amounts to about 10 events
above 2.6 MeV, and causes a small shift in the $\delta m^2$ value\cite{inoue}.The additional background events below 2.6 MeV make it difficult to extract a
geo-neutrino signal at present.  Kamland presented results on spectrum
distortion and an "L/E" analysis which showed the characteristic 
oscillatory pattern.  Kamland collaboration is now working on further 
purification with $^7Be$ solar $\nu_e$ detection in mind, with a possible 
2007 date\cite{inoue}.

\section{Terra Incognita I}
The idea behind the geo-neutrino measurement is to determine the
distribution of $U$ and Th in the crust and the mantle, to confirm that
40\% of the earth's heat comes from is radioactivity\cite{fiorentini}.  
When Kamland lowers
the background, it should be able to have the first results on
geo-neutrinos.  However, at Kamioka as well as at Gran Sasso (site of
Borexino), the reactor background is quite high\cite{calaprice}:

\begin{tabular}{ccc}
  Site    & Geo-$\nu$ Rate    & Reactor BG  \\ \hline
  Kamioka &  5.4             &  1.5  \\
  GSL     &  5.9             &  0.65  \\
  SNO     &  6.8             &  1.3 \\
  Hawaii  &  3               &  0.027  \\
  Tibet   &  7.2             &  0.054  \\
\end{tabular}

It is clear that a 1KT Kamland style detector near Hawaii would be most
suitable for a clean accurate measurement of geo-neutrinos, at least 
for the flux from the mantle\cite{fiorentini}.

The possibility of a Geo-reactor at the center of the earth's core has
been raised by Herndon\cite{herndon}.  This would be a breeder fission reactor operating at a power of about 3-10TW.  As the earth cooled;  U, Th in 
alloys and
sulphides could have sunk to the centre and formed an OKLO-style reactor.  
This would also explain the low amount of oxygen and the high amount of
$^3He$ and can provide a fluctuating energy source for geomagnetism.  
This alternative to the dynamo mechanism is not yet widely accepted, but is
beginning to be taken seriously.  In any case, observation of
$\bar{\nu}_e's$ from the reactor (or non-observation) would be a crucial
test\cite{raghavan}.  The Kamland-in-Hawaii would be 
an ideal detector for this purpose as well.

In her review, Turck-Chieze\cite{turcke} emphasized that there is no unified view of
stars yet, and that three dimensional calculations of stellar evolution are
still in future.  As for the solar neutrino fluxes, the CNO fluxes are 
now expected to be somewhat lower.  There seems to be a need 
for some new ingredients
beyond SM to describe the sun fully.

The desiderata for future solar neutrino experiments seem to 
be\cite{nakahata}: (a) to see
the up-turn at low energy in SNO, (b) to see the day-night asymmetry
(2-3\% in SNO, 1-2\% in SK), (c) extract the CNO flux, (d) proof of LMA,
e.g. by confirming $^7Be$ flux (Kamland and Borexino), (e) precision CPT
test by Kamland-solar comparison, (f) possible new $\nu$-physics and 
(g) measurement of the ratio of neutrino luminosity to that in photons:
$L_\gamma /L_\nu$.

Several future solar neutrino programs were discussed\cite{nakahata}.  
Borexino and SNO
(LS) will be able to measure the {\it pep} line as well CNO flux.  The $^{11}C$
background studies which aim to reduce it are crucial for this\cite{Franco}.  LENS aims to measure pp $\nu_e's$ via charged current interaction
in real time.  XMASS with 10T of Xe is an ambitious project and over 5 years
can measure pp $\nu_e's$ in neutrino-electron scattering and measure $\sin^2 \theta_{12}$ at 1\% level.  
Future Megaton $H_2 O$ \^{c} facilities\cite{goodman} can measure $^3He-p$ as well as
day-night asymmetry accurately.  Intermediate base line (70 km) reactor
experiments can also measure $\sin^2 \theta_{12}$ to 3\%, and help remove
degeneracies in the neutrino parameters\cite{minakata}.

\section{Terra Cognita II}
In atmospheric neutrinos, there was real progress in modeling 
fluxes\cite{stanev}.  
New flux calculations including 3-dimensional codes represent the state of the
art, and have reached a new level of sophistication. 
The main (and irreducible) uncertaintly affecting the extraction of 
oscillation parameters is now the uncertainty in the knowledge of primary 
fluxes.
Experimentally the two main things are the observation of unambigous 
signature of oscillations in the dip in L/E and the laboratory 
(K2K) confirmation of the atmospheric deficit\cite{kajita}.  
I was  most pleased to see the confirmation of the oscillatory behaviour,
something that we have been waiting for since 1988! 

We heard about the Long Baseline and neutrino factory plans for the
future.  In Japan, there is a well-developed program\cite{kajita} 
with T2K already underway\cite{kaneyuki} and planning going 
ahead for phase II.  For T2K initially
the beam will come from 0.75 MW proton beam at 50 GeV from JPARC to SuperK, with 295 km baseline.
The main goal will be to  measure $\theta_{23}$ and $\delta m^2_{32}$ with a
higher precision,  and search for a non-zero $U_{e3}$.  
In  phase II with 4 MW power and a new MegaT detector, the goal will be to 
search for CP violation, which would be observable
for $\delta > 20^0$ and $\sin^2 2\theta_{13} > 0.01$.

In the U.S.\cite{grzelak} MINOS will start taking data in 2005, and reach 10\%
accuracy in $\delta m^2_{23}$ and be able to test CPT because of the ability
to separate $\nu$ and $\bar{\nu}$ events(with the atmospheric 
$\nu's)$. Post-MINOS plans involve NOVA which can probe $\theta_{13}$ at a level
of 0.01 (in $\sin^2 2 \theta_{13})$, 
measure $\sin^2 2  \theta_{23}$ at  a 1-2\%
level, measure the sign of $\delta m_{23}^2$ and probe CP violation with 
the proton driver.  The off-axis detector for NOVA should be as large as 
possible.  The precise nature of the detector at a baseline of about 800Km is under discussion.

In Europe, the LBL program\cite{ronga} with CNGS is proceeding on schedule,
with first beam expected in May 2006.   Construction of OPERA,  
with a reasonable sensitivity for $\nu_\tau$ events, is also underway.  
ICARUS will use a different technique for $\nu_\tau$ events.  Another proposal is to use the CERN-TARANTO(CNFT) baseline  to utilize the second maximum and reach $\sin^2 2 \theta_{13}$ to 0.002 level.  There was some discussion of using beta-beams with CERN-Frejus(or GSL) baseline  and even using LHC to make high energy beta-beams\cite{gomez}.

\section{Terra Incognita II}
Absolute neutrino masses are probed in (i) end point measurements
in (e.g. tritium) beta decay\cite{drexlin}, (ii) neutrinoless 
double beta decay\cite{klapdor} and 
(iii) large scale structure in cosmology\cite{hannestad}.  The three are sensitive to
different combinations:
\begin{eqnarray}
\mbox{Tritium}:  \quad  \quad  m_\beta & = & \left [ \sum_{i} \mid 
U_{ei} \mid^2 m_i^2 \right ]^{1/2}\\ \nonumber
\beta\beta 0 \nu:  \quad  \quad   m_{\beta \beta}& = & \sum_{i} 
U_{ei}^2 m_i \\ \nonumber
LSS:   \quad \quad \Sigma & = & \sum_{i} m_i
\end{eqnarray}

If we neglect, for simplicity, $\mid U_{e3} \mid^2$ then
\begin{eqnarray}
m_\beta & \cong & \left [ m_1^2 + s^2_{12} \delta m_{21}^2 \right ]^{1/2}
\\ \nonumber
m_{\beta \beta} & \cong & 
\left \{
\begin{array}{c}
m_1 + s^2_{12} \sqrt{\delta m_{21}^2} \approx m_1 \\
\frac{1}{2} m_1 - s^2_{12} \sqrt{\delta m_{21}^2} \approx \frac{1}{2} m_1
\end{array} \right. \\ \nonumber
\Sigma & \cong & 3m_1 + \sqrt{\delta m_{21}^2} \pm \sqrt{\delta m_{32}^2}
\end{eqnarray}
where the two cases for $m_{\beta \beta}$ correspond to constructive or
destructive Majorana phases and  the two signs in $\Sigma$ corresponds to normal
and inverted hierarchy\cite{ferruglio}.  
If we consider the range of values implied by the
results for $m_{\beta \beta}$ that have been reported\cite{klapdor} 
(but not yet
confirmed), and take a value in the middle of
this range,
\begin{equation}
m_{\beta \beta} \cong 0.5 eV
\end{equation}

then the value for $\Sigma$ is 1.5 eV for constructive phase and 3 eV for
destructive phase.  But a value for $\Sigma$ as high as 3eV is disfavored
by the recent cosmological data\cite{hannestad}. Hence a constructive 
phase is favored, 
quasi-degeneracy is also favored and a value for $m_1 \sim m_\beta \sim 
0(eV)$ is expected.  The
next round of tritium experiments (e.g. KATRIN)\cite{drexlin} 
should be able to confirm
this.  This is true for any value of $m_{\beta \beta}$ larger than about 0.25
eV.  The current constraints may tighten and make life very interesting.  We
await results from future double beta decay experiments\cite{brofferio}, 
KATRIN and
further refinement and improvement of cosmological bounds. The extraction of
the effective mass from neutrinoless double beta decay rates depends
crucially 
on the improvements in the knowledge of the nuclear matrix
elements\cite{faessler}. 
Needless to say,
neutrinoless double beta decay is very important to establish the Majorana
natures of the neutrino; unfortunately a null result does not establish 
a Dirac nature. It is quite
remarkable that explanations of baryon asymmetry based on leptogenesis
place bounds on neutrino masses in the (0.15 to 1) eV range\cite{bari}.

\section{Terra Incognita III(?)}
There was a lively discussion of theory issues during the 
meeting\cite{ferruglio,senjanovic}.  In my
view, while there have been many interesting proposals, we are still
waiting for a breakthrough.  This is true for not only neutrino mass
matrix, but flavors in general.  
We would like to understand the smallness of neutrino masses, the near
maximality of some mixing, the pattern of the masses, the near degeneracy
if true.  The presence of broken family symmetries is still an open
question.  Do GUTS play a role (any) in the mass and mixing patterns?
Ferruglio\cite{ferruglio} gave an elegant summary of the current status of theory.

What about making predictions?  Recently there have been many attempts at
predicting the remaining mixing element: $U_{e3}$.  The prediction run the
gamut from the maximum allowed value all the  way to nearly zero\cite{ferruglio}.  
There is an interesting proposal that I learned
from Bjorken and from Javier Ferrandis\cite{ferrandis}.  
The idea is the following.  Suppose
that the neutrino mass matrix is diagonalized by a matrix $U_\nu$ which
has the form:
\begin{eqnarray}
U_\nu \cong
\left (
\begin{array}{ccc}
\cdot & \cdot & 0 \\
\cdot & \cdot & 1/\sqrt{2} \\
\cdot & \cdot & 1/\sqrt{2}
\end{array} \right )
\end{eqnarray}

The charged lepton mass matrix has the form analogous to the down quark
mass matrix (which leads to a successful prediction for $\theta_c$ of
$\theta_c \approx \sqrt{m_d/m_s})$:
\begin{eqnarray}
M_\ell \sim \left (
\begin{array}{ccc}
0 & a & \cdot \\
a & b & \cdot \\
\cdot & \cdot & \cdot
\end{array} \right )
\end{eqnarray}

In this case $\theta_{12}^\ell \sim \sqrt{\frac{me}{2m_{\mu}}},$ and the
full
\begin{equation}
U_{MNSP} = U_\ell^+ U_\nu
\end{equation}
has now a $U_{e3} \sim \sqrt{\frac{me}{2m_{\mu}}} \sim 0.052$.
This is the resulting prediction for $U_{e3}$ and it is testable in
future ambitious reactor proposals we heard about here\cite{several}.

Recently there has been some discussion of the so-called QLC 
relation\cite{smirnov}:
\begin{equation}
\theta^{12}_{CKM} + \theta^{12}_{MNSP} = \theta_c + \theta_{solar} \approx 45^0 
\end{equation}
where $\theta_c \sim 13^0$ and $\theta_{solar} = 33 \pm 2^0$.  In fact there 
is also the approximate relation:
\begin{equation} 
\theta^{23}_{CKM} + \theta ^{23}_{MNSP} \simeq 45^0
\end{equation}
within errors\cite{PDG}.  We  have no idea if these are accidents or genuine 
hints with some deeper meaning.  A generalised QLC relation is then suggested:
$U_{CKM} \times U_{MNS} = U_{BM}$. This leads me to a third relation:
\begin{equation} 
\theta^{13}_{CKM} + \theta ^{13}_{MNSP} \simeq 0
\end{equation}
If the RHS above is really close to zero,  the  implication would be that
\begin{equation}
\mid U_{e3} \mid = \mid V_{ub} \mid \approx 0.004,
\end{equation}
in which case  $\mid U_{e3} \mid$ is out of reach of 
experiments\cite{several}.  I hope
this is not the case.

\section{Atlas Coelestis: Neutrinos from Heavens}
I would like to discuss in the rest of the talk two topics which were not
discussed at great length
during the meeting.  One is the uses of high energy
astrophysical neutrinos  and the other is detection of relic neutrino
background. While the production\cite{semikoz} of and detection\cite{halzen} of high energy astrophysical
neutrinos was discussed, all the various uses they can be put to was not
discussed in any detail. Likewise, while relic neutrinos were discussed
their possible detection was not discussed. I review these 
topics very briefly.

We make two basic, reasonable assumptions:  
one is that distant high energy neutrino sources exist; 
and two that in the near future, very large volume, well 
instrumented detectors of sizes of order of KM3 
(as discussed here)\cite{halzen} and beyond will be operating.

From these sources, we expect half as many 
$\nu_e's$ as $\nu_\mu's$ and virtually no $\nu_\tau's$.  
This comes about simply 
because the neutrinos are thought to originate in 
decays of pions (and kaons) and subsequent decays 
of muons. The flux ratio of $\nu_e: 
\nu_\mu: \nu_\tau = 1:2:0$ is canonical for most sources\cite{semikoz}.

With the current knowledge of neutrino masses and mixings 
as summarized earlier and with $\delta m^2 L/4E$ so large
that the oscillations are always averaging out, a flux ratio of 
$\nu_e: \nu_\mu: \nu_\tau = 1:2:0$ gets converted into one of 1: 1: 1. 
Hence the flavor mix expected at arrival 
is simply an equal mixture of $\nu_e, \nu_\mu$ and 
$\nu_\tau$ as was observed long ago\cite{learned2,athar}.

If this universal flavor mix is confirmed by future observations, our
current knowledge of neutrino masses and mixings is reinforced and
conventional wisdom about the beam dump nature of the production process
is confirmed as well.  However, it would much more exciting to find
deviations from it, and learn something new.  How can this come about?  
There are quite a few ways in which this can happen.  Below is a shopping
list of a variety of ways in which this could come to pass.

The first and simplest is that initial flavor mix is NOT 1 : 2 : 0.  This
can happen when there are strong magnetic fields causing muons 
to lose energy
before they decay, and there exist models for neutrino 
production in AGN's in which this does happen\cite{rachen}.  In this case 
the $\nu_e's$ have 
much lower energies
compared to $\nu'_\mu s$ and effectively the initial 
flavor mix is 0 : 1 : 0 and averaged out oscillations 
convert this into 1/2 : 1: 1 on arrival.

The possibility that the mass differences between neutrino mass
eigenstates are zero in vacuum (and become non-zero only in the presence
of matter) has been raised recently\cite{kaplan}.  If this is true, then the final
flavor mix should be the same as initial namely: 1 : 2 : 0.

Neutrino decay is another important possible way for the flavor mix to
deviate significantly from the democratic mix\cite{beacom1}.  If neutrinos decay, then
in general, the heavier neutrinos are expected to decay into the lighter
ones via flavor changing processes.  It has been shown that the only
possible interesting modes are two body modes into a lighter neutrino and
massless boson\cite{pakvasa1}.  With neutrinos from these sources we can probe lifetimes
many orders of magnitude longer than the current bounds. Relic supernova
signals of $\bar{\nu}_e's$ can probe even longer lifetimes\cite{montanino}.

For normal hierarchy in which both $\nu_3$ and $\nu_2$ decay, only the
lightest stable eigenstate $\nu_1$ survives.  In this case the flavor
ratio is $U^2_{e1} : U^2_{\mu 1} : U^2_{\tau 1}$\cite{pakvasa2}.  Thus if $U_{e3} = 0,$
then
\begin{equation}
\phi_{\nu e} : \phi_{\nu \mu} : \phi_{\nu \tau} \simeq 5 : 1 : 1,
\end{equation}
for the neutrino mixing parameters given above.  This is an extreme
deviation of the flavor ratio from that in the absence of decays and
should be easy to distinguish.  For inverted hierarchy, $\nu_3$ is the
lightest and hence stable state, and so\cite{beacom1}
\begin{equation}
\phi_{\nu_{e}} :  \phi_{\nu_{\mu}} : \phi_{\nu _{\tau}} = U^2_{e3} : 
U^2_{\mu 3} : U^2_{\tau 3} = 0 : 1 : 1.
\end{equation}

When $U_{e3}$ is not zero, and the hierarchy is normal, it is 
possible to obtain information on the values of $U_{e3}$ 
as well as the CPV phase $\delta$\cite{beacom2}.  
The flavor ratio $e/\mu$ varies from 5 to 15 (as $U_{e3}$ goes 
from 0 to 0.2) for $\cos \delta =+1$ but from 5 to 2 for $\cos 
\delta =-1$.  The ratio $\tau/\mu$ varies from 1 to 5 $(\cos 
\delta = +1)$ or 1 to 0.2 $(\cos \delta =-1)$ for the 
same range of $U_{e3}$.

If the decays are not complete and if the daughter does 
not carry the full energy of the parent neutrino; the resulting 
flavor mix is somewhat different but in any case it is still quite 
distinct from the simple $1:1:1$ mix\cite{beacom1}.

If neutrinos have flavor (and equivalence principle) violating couplings
to gravity (FVG), or Lorentz invariance violating couplings; then there
can be resonance effects which make for one way transitions (analogues of
MSW transitions) e.g. $\nu_\mu \rightarrow \nu_\tau$ but not vice versa.  
In case of FVG for example, this can give rise to an anisotropic deviation
of the $\nu_\mu/\nu_\tau$ ratio from 1, becoming less than 1 for events
coming from the direction towards the Great Attractor, while remaining 1
in other directions\cite{minakata2}.

Another possibility that can give rise to deviations of the flavor mix
from the canonical 1 : 1 : 1 is the idea of neutrinos of varying mass
(MaVaNs), in which neutrino masses are related to the dark energy\cite{nelson}.  The
end result is that the neutrino mass depends inversely on neutrino
density, and hence on the epoch.  As a result, if the sterile neutrino
mixes with a flavor neutrino, the mass difference varies along the path,
with potential resonance enhancement of the transition probability into
the sterile neutrino, and thus change the flavor mix.  For example, if one
resonance is crossed enroute, it can lead to a conversion of the lightest
(mostly) flavor state into the (mostly) sterile state, thus changing the
flavor mix to $1-U^2_{e1} : 1 - U^2_{\mu 1} : 1- U^2_{\tau 1} \approx 1/3$
: 1 : 1, in case of inverted hierarchy and similarly $\approx$ 2 : 1 : 1
in case of normal hierarchy\cite{hung}.

If each of the three neutrino mass eigenstates is actually a doublet with
very small mass difference (smaller than $10^{-6} eV)$, then there are no
current experiments that could have detected this.  It turns out that the
only way to detect such small mass differences $(10^{-12} eV^2 > \delta
m^2 > 10^{-18} eV^2)$ is by measuring flavor mixes of the high energy
neutrinos from cosmic sources\cite{beacom3}.

The flavors deviate from the democratic value of $\frac{1}{3}$ by
\begin{eqnarray}
\delta P_e & = & - \frac{1}{3}
\left [ \frac{3}{4} \chi_1 + \frac{3}{4} \chi_2 \right ], \\ \nonumber
\delta P_\mu = \delta P_\tau & = & - \frac{1}{3} 
\left [ \frac{1}{8} \chi_1 + \frac{3}{8} \chi_2 + \frac{1}{2} \chi_3\right ]
\end{eqnarray}
where $\chi_1 = \sin^2 (\delta m^2_i L /4E).$  The flavor ratios deviate
from 1 : 1 : 1 when one or two of the pseudo-Dirac oscillation modes is
accessible.  In the ultimate limit where $L/E$ is so 
large that all three oscillating factors have averaged 
to $\frac{1}{2}$, the flavor ratios return to 1 : 1 : 1, 
with only a net suppression of the measurable flux, by a factor of 1/2.

To summarize, the measurement of neutrino flavor mix at neutrino
telescopes is absolutely essential to uncover new and interesting physics
of neutrinos\cite{beacom4}. In any case, it should be evident that the construction of
very large neutrino detectors is a ``no lose'' proposition.

\section{Atlas Coelestis: Relic Neutrino Detection} 
Turning to the relic neutrinos, we know\cite{kolb} that the effective 
temperature ``$T_\nu$'', today is about $1.9^0K \sim 1.7X10^{-4}$ eV. 
The number density $(\nu+ \bar{\nu})$ is about 115/cc.  (This assumes 
that we do not live in a ``neutrino-free'' 
universe!\cite{glashow}).  How can we detect
these neutrinos?

The average momentum of relic neutrinos is $3.2T_\nu \sim 5.2 \times
10^{-4} eV/c$.  The neutrino current density is $cn_\nu \sim 10^{13} 
cm^{-2} s^{-1}$ for massless neutrinos and $5.10^9 cm^{-2} s^{-1}$ 
for mass of O(eV). The effective interaction Hamiltonian for neutrinos with
neutral matter is proportional to $a_e = (3Z-A)$ for $\nu_e$ and $a_\mu =
(A-Z)$ for $\nu_\mu$ and $\nu_\tau$.  The $\nu_\alpha$-scattering cross
section (at very low energies) on nuclei then goes as
\begin{equation}
\sigma_\alpha \sim \frac{a^2_\alpha \ G^2_F \ m^2_\nu}{\pi}
\label{goes}
\end{equation}
Many early proposals to detect relic neutrinos by reflection or coherent
effects turned out to be incorrect.  There are three methods which some
day may prove to be practical.

The first is a proposal due to Stodolsky.  The idea needs 
neutrino degeneracy i.e. excess of $\nu ($or $\bar{\nu})$ over $\bar{\nu}$
(or $\nu)$ to work.  Then a polarized electron moving in a background of
CMB neutrinos can change its polarization due to the axial 
vector parity violating interaction.  The effective neutrino density (for
$n_\nu \gg n_{\bar{\nu}})$ goes as $p^2_f/6 \pi^2$ where $p_f$ is the Fermi
momentum.  The effective interaction goes as 
\begin{equation}
H_{eff} \sim \frac{2 G_F}{\sqrt{2}} \ \overline{\sigma}_e \cdot 
\overline{v} \ n_\nu
\end{equation}
 
With $v \sim 300km/sec$ and $p_f \sim O(eV)$ this leads to a rotation of
the polarization of about 0.02'' in a year.  Can such small spin rotations
can be detected?  Certainly not at present, but technology may someday
allow this.

For a magnetized macroscopic object, this interaction can also give rise
to a torque and lead to an acceleration which can be estimated as
\begin{equation}
a \sim 10^{-27} cm/sec^2
\end{equation}
for some typical dimensions\cite{gelmini}.  
Several proposals have been made but all seem
to need future technological breakthrough\cite{hagmann}.

The second method is one suggested by Zeldovich and 
collaborators\cite{shvartsman}.  The idea is to take advantage of momentum 
transfer in neutrino-nucleus scattering.  
Consider an object made up of small spheres of radius a 
$\approx \lambda$ (neutrino wavelength) packed loosely 
with pore sizes also of the same size (to
avoid destructive interference).  If the number of atoms 
in the target is $N_A$ then the effective coherent cross-section is
\begin{equation}
\sigma = \sigma_\alpha \ N^2_A
\end{equation}
where $\sigma_\alpha$ is as given in Eq.~\ref{goes}.  Assuming 
total reflection, momentum transfer is
\begin{equation}
\Delta p \cong 2m_\nu \ v_\nu
\end{equation}
and the force $f = j_{\nu} \sigma \Delta p$ is given by
\begin{equation}
f=2n_{\nu} \sigma_{\alpha} \ N^{2}_{A} m_{\nu} v^2_\nu
\end{equation}
The most optimistic estimates are obtained by 
assuming some clustering
$(n_\nu \sim 10^7/c.c.), m_\nu \sim 0(eV), 
v_\nu \sim 10^7 cm/s, \rho \sim 10gm/cc$; leading to
\begin{equation}
a= \frac{f}{m} \ = \frac{f}{N_A m_N} \ \sim 10^{-23}(a_\alpha/A)^2cm.s^{-2}
\end{equation}  

Such accelerations are at least ten orders of magnitude removed from
current sensibility and possible detection remains far in future\cite
{hagmann2}.  In
addition, this effect is absent for Majorana neutrinos or suppressed very
much\cite{gelmini}.

The third possibility is based on the  proposal by Weiler in 
1982\cite{weiler}.  The basic
idea is as follows.  If neutrinos have masses in the eV range and there
are sources of very high energy neutrinos at large distances, then the
they  can annihilate on the relic  $\bar{\nu}$ and make a $Z^0$
on-shell at resonance creating an absorption dip in the neutrino spectrum.  
The threshold for $Z$ production would be at $E \sim m^2_Z/2m_\nu$ which
is about $4.10^{21}$ eV for $m_\nu \sim 0(eV)$.  
This seemed like an unlikely possibility, since
it required large neutrino fluxes at very high energies to see the
neutrino spectrum and then the absorption dip.  But all this changed
dramatically recently with the hints of a  
signal of cosmic rays beyond the GZK
cut-off.  The GZK cut-off is the energy at which cosmic ray protons pass
the threshold for pion production off the CMB photons. This is at an
energy $E \sim m_\pi m_p/E_\gamma \sim 6.10^{19}$ eV.  Above this energy,
the mean free path of protons is less than 100MPc and hence these protons
have to  originate ``locally''. The flux for higher energies should then decrease dramatically since we
believe the cosmic rays are not produced locally. Recently, what used to
be hints of the cosmic ray signal extending beyond this cut-off, may have
become a clear signal.  The events are most likely due to protons.  
Then an explanation is called for.  One intriguing proposal is that these
events are nothing but a signal for the $Z's$ produced by the $\nu
\bar{\nu} \rightarrow Z$ process with the protons coming from the
subsequent Z decay\cite{weiler1}! 
Of course, the original problem of needing sources of
high energy neutrinos remains.  If this explanation is valid, we may have
already seen (indirect) evidence for the existence of relic neutrinos.  
In principle, this proposal can be tested: (i) the events should point back
out at the neutrino sources; (ii) there is an eventual cut-off when the
energy reaches the threshold energy for Z production, $E \sim 4.10^{21}
\left (\frac{eV}{m_\nu} \right )$ eV; (iii) $\gamma/p$ ratio should be
large near threshold and (iv) the large $\nu$-flux should be eventually
seen directly in large $\nu$-telescopes. This pretty picture may be on
the verge of being ruled out by the ever tightening bounds on flux of diffuse
high energy neutrinos\cite{gorham}.
                   I am afraid it may be a very long time before we have 
clear cut, unambiguous direct detection of relic neutrinos.

\section{Acknowledgements}
                I would like to thank the organisers for the outstanding
hospitality and the stimulating atmosphere of the workshop.  I look forward 
eagerly to the next NOW workshop two years from now! This work was
supported in part by U. S. D. O. E. under grant DE-FG02-04ER41291.

\end{document}